# Narrowband terahertz radiation by an impulsive stimulated Raman scattering in an above-room-temperature organic ferroelectric benzimidazole

M. Sotome[1], N. Kida[1,*], S. Horiuchi[2], and H. Okamoto[1,3]

[1]*Department of Advanced Materials Science, The University of Tokyo, 5-1-5 Kashiwa-no-ha, Chiba 277-8561, Japan*

[2]*National Institute of Advanced Industrial Science and Technology (AIST), Tsukuba 305-8562, Japan*

[3]*AIST-U. Tokyo Advanced Operando-Measurement Technology Open Innovation Laboratory, National Institute of Advanced Industrial Science and Technology, Chiba 277-8568, Japan*

E-mail: kida@k.u-tokyo.ac.jp

**Abstract:**

We observe a terahertz radiation from a hydrogen-bonded organic molecular ferroelectric 5,6-dichloro-2-methylbenzimidazole excited by a femtosecond laser pulse at room temperature. The emitted terahertz wave consists of three oscillatory components, the frequencies of which agree with those of Raman- and infrared-active phonon modes. This suggests that the terahertz radiation is attributed to polarization modulations by infrared-active phonons excited via the impulsive stimulated Raman scattering processes. By taking into account the Raman polarizability tensor and dipole-moment for each phonon, we succeeded in reproducing not only the spectrum of the terahertz radiation, but also its time characteristic. The analysis method is discussed in detail. Our result provides a new way for the light-induced terahertz radiation in organic ferroelectrics.

The second-order optical nonlinearity in noncentrosymmetric media is useful for the frequency conversion of lights, which include not only the sum and difference frequency generation but also the optical parametric effect [1,2]. In this context, ferroelectric materials are being intensively studied [1]. In oxide ferroelectrics such as $LiNbO_3$ and $KH_2PO_4$, highly efficient second-order optical nonlinearity has been reported, which is enough for practical use as nonlinear optical materials. Explorations of organic ferroelectrics are also important, since organic materials have advantages due to their low cost and environmentally benign characteristics. However, it had been reported that their Curie temperatures are much lower than room temperature. Recently, above-room-temperature ferroelectricity has been found in hydrogen-bonded molecular crystals having π-electron systems [3] such as 4,5-dihydroxy-4-cyclopentence-1,2,3-trione (croconic acid) [4] and 2-phenylmalondialdehyde [5]. In these materials, cooperative proton displacements and asymmetric π-electron configurations along the hydrogen-bonded direction cause the ferroelectric polarization.

In croconic acid and 2-phenylmalondialdehyde, we have recently demonstrated that terahertz electromagnetic wave can be radiated by an irradiation of a femtosecond laser pulse [6,7] and that the terahertz-radiation can be ascribed to a difference frequency generation (DFG) within an incident laser pulse, which is described by the second-order nonlinear optical susceptibility $\chi^{(2)}$. This process is sometimes called optical rectification (OR) [8] and is recognized as a typical terahertz-radiation mechanism in various noncentrosymmetric media such as ZnTe [9] and 4-N,N-dimethylamino-4'-N'-methyl stilbazolium tosylate (DAST) [10].

In the present study, we report that effective terahertz radiation is possible via a different mechanism in a hydrogen-bonded organic molecular ferroelectric, 5,6-dichloro-2-methylbenzimidazole (DCMBI) [11]. By the irradiation of a femtosecond laser pulse on a single crystal of DCMBI, we observe an emission of a terahertz wave consisting of several oscillatory components, which cannot be attributed to the OR mechanism. The central frequencies of those oscillations correspond to three phonon modes, which are both Raman- and infrared-active, indicating that the terahertz radiation originates from the infrared-active phonons excited via the impulsive stimulated Raman scattering processes. By taking into account the polarizabilities of Raman scatterings and the dipole moments of infrared absorptions of phonons, we show that both the spectrum and time characteristic of the terahertz radiation are well simulated.

The crystal system of DCMBI is an orthorhombic with a space group of $Pca2_1$ (point group of $mm2$), as illustrated in Fig. 1(a) [11]. The imidazole molecules with a large dipole-moment are connected by intermolecular hydrogen bonds along the $c$-axis, as indicated by red lines. The protons are placed at off-center positions with asymmetric π-electron distribution, resulting in a large spontaneous polarization $P$ (~10 $\mu C/cm^2$) along the hydrogen-nitrogen-bond direction, i.e., $c$-axis [11]. The polarization reversal is possible by cooperative proton transfers and π-bond switching.

Single crystals of DCMBI were grown by a previously reported method [11]. For the steady-state reflectivity and transmission measurements in the visible to mid-infrared region, we used a Fourier-transformed infrared spectrometer (FTIR) and a grating spectrometer, both of which are equipped with specially designed optical microscopes. For the transmission measurement in the terahertz region, we adopted a terahertz time-domain spectroscopy (TDS) [9], in which a mode-locked Ti:sapphire laser with the central wavelength of 800 nm (1.55 eV), the repetition rate of 80 MHz, and the pulse width of 100 fs was used as a light source. A (110)-oriented ZnTe emitter with the thickness of 2 mm excited by a Ti: sapphire laser pulse was used as a light source and a photoconductive antenna with low-temperature-grown GaAs (LT-GaAs) excited by a part of the Ti: sapphire laser pulse was used as a detector. For the Raman scattering experiments, we used a Raman spectrometer equipped with a He-Ne laser (632.8 nm) as the light source and an optical microscope. In the terahertz radiation experiments, the output pulse of the mode-locked Ti: sapphire laser mentioned above was used as an excitation and the electric-field waveform of the terahertz radiation was also detected by a photoconductive antenna. The detection range is 0.5 to 3 THz. All the optical measurements were carried out at 294 K.

First, we show the fundamental optical properties from the visible to mid-infrared region. Figure 1(b) shows the polarized reflectivity $R$ and transmittance $T$ spectra, which are measured on a single crystal with the thickness $d$ of 150 μm for the electric fields of light $E^\omega$ parallel to the $b$-axis ($E^\omega||b$) and to the $c$-axis ($E^\omega||c$). A sharp decrease of $T$ above 3.8 eV for $E^\omega||b$ is attributable to the π-π* transition peaked at 4.1 eV, which is clearly observed in the polarized reflectivity spectrum along the $b$-axis. $T$ also decreases at almost the same position for $E^\omega||c$, suggesting that the π-π* transition has a finite component along the $c$-axis. Spiky structures below 0.5 eV are due to optical phonons. We calculated the spectra of the refractive index $n$ from the $R$ spectra by the Kramars-Kronig transformation and the spectra of the absorption coefficient $\alpha$ from the $R$ and $T$ spectra using the relationship, $\alpha = -\frac{1}{d}\ln\left(\frac{T}{(1-R)^2}\right)$. The obtained $n$ and $\alpha$ spectra are shown in Figs. 1(c) and 1(d), respectively. From 1 to 3 eV, no optical anisotropy is observed and $n$ is nearly constant (~1.8) and $\alpha$ is negligibly small (< 6 cm$^{-1}$). We reproduce $n$ spectra by using Sellmeier relation and estimate the group refractive index $n_g$ spectra, which are shown by dotted lines in Fig. 1(c). $n_g$ at 1.55 eV (indicated by vertical arrows) for $E^\omega||b$ and $E^\omega||c$ are 1.88 and 1.83, respectively.

Next, we show the results of terahertz-radiation by an irradiation of a femtosecond laser pulse. The experimental set-up in the laboratory coordinate *XYZ* is shown in Fig. 2(a). The experiment was performed on the same single crystal as used in the optical spectroscopy mentioned above. The LT-GaAs detector is set so that the *X*-axis component of the electric field $E_X(t)$ is selectively detected. Figures 2(b) and (c) show the time characteristics of $E_X(t)$ for the electric fields $E^\omega$ of the incident laser pulse parallel to the *c*- and *b*-axes ($E^\omega||c$ and $E^\omega||b$), respectively. The time origin is arbitrary. For $E^\omega||c$, we can see a nearly single-cycle electric field around 0 ps, which is followed by an oscillatory component at least up to 8 ps. In contrast, for $E^\omega||b$, we observe a prominent oscillation signal, which seems to consist of several oscillatory components with different frequencies.

In organic ferroelectrics, domains with different polarization directions sometimes coexist in an as-grown single crystal [6,12]. To investigate a domain structure, we measured $E_{\mathrm{THz}}(0)$ at various positions of the DCMBI crystal using a raster scan and obtained the domain image over the crystal [6,12-15]. The effectiveness of this domain-imaging method was demonstrated in other ferroelectrics such as croconic acid [6,12-14]. The result of DCMBI revealed that the crystal consists of a single domain state (Supplemental Material S1 [16]) and a femtosecond laser pulse with a spot diameter of ~25 μm is irradiated within a single-domain state.

Next, we discuss the second-order nonlinear optical process related to the terahertz radiation (Supplemental Material S2 [16]). When the second-order nonlinear optical process is dominant, $E_X(\omega)$ is expressed as

$$E_X(\omega) \propto (d_{32}\sin^2\theta + d_{33}\cos^2\theta)I, \quad (3)$$

where *I* is the laser intensity and $\theta$ is the light-polarization angle relative to the *X*-axis [Fig. 2(a)]. Figure 2(d) shows $\theta$ dependence of $E_X(\omega)$ integrating in the frequency range of 0.8-1.2 THz and 1.8-2.2 THz, which are indicated by green and pink circles, respectively. Equation (3) well reproduces $\theta$ dependence of $E_X(\omega)$ as shown by the solid lines; the ratio $(d_{32}/d_{33})$ are evaluated to be 0.54 and 1.54 by integrating in the frequency range of 0.8-1.2 THz and 1.8-2.2 THz, respectively. As shown in Fig. 2(e), the terahertz radiation intensities are proportional to $I$ as expected from Eq. (3). These results demonstrate that second-order nonlinear optical effect dominates the light-induced terahertz radiation from crystalline DCMBI.

In the terahertz radiation experiments, it is important to consider the phase-matching condition between an incident laser pulse and a generated terahertz wave. We calculate the coherence length $l_c (= \frac{\lambda}{2} \frac{1}{|n_g - n_{THz}|})$ along the $b$- and $c$-axes, where $\lambda$ and $n_{THz}$ are the wavelength and refractive index of the emitted terahertz waves, respectively. From the polarized $T$ measurements of a 397 μm-thick $a$-axis-oriented crystal by terahertz TDS, we estimated $n_{THz}$ spectra (Supplemental Material S3 [16]). Using the values of $n_g$=1.83 (1.88) at 1.55 eV [Fig. 1(c)] and $n_{THz}$=1.68-1.86 (1.68-1.87) at 0.5-3 THz for $E^\omega || c$ ($E^\omega || b$), we obtain the frequency dependence of $l_c$; $l_c$=2.96 mm at 1 THz (1.34 mm) for $E^\omega || c$ ($E^\omega || b$) (Supplemental Material S3 [16]). As shown in Fig. S2(c), $l_c$ shows the sharp increase, which originates from a good phase-matching condition ($n_g \sim n_{THz}$) due to the presence of the sharp dispersion of the phonon modes [17]. However, in both configurations, $l_c$ at 0.5-3 THz is much longer than the sample thickness (150 μm), indicating the good phase-matching condition in terahertz radiation experiments. This also excludes the generation of the narrowband terahertz radiation in DCMBI due to the resonant enhancement of $l_c$. In addition, the penetration depths ($1/\alpha$ =1.6 mm for $E^\omega || c$ and 1.0 mm for $E^\omega || b$) [Fig. 1(d)] of the incident femtosecond laser pulse also far exceed the sample thickness (150 μm), so that terahertz waves are considered to be emitted from all of the DCMBI crystal. The terahertz radiation efficiency is comparatively high for both $E^\omega||c$ and $E^\omega||b$ and is estimated to be ~24% and 17% of that in a typical terahertz emitter of (110)-oriented ZnTe.

To investigate the nature of the observed terahertz radiations, we performed the Fourier transformation of the obtained $E_X(t)$ profiles, which are shown by circles in Fig. 3(a). The frequency-resolution (0.05 THz) is indicated by a pair of vertical lines. For $E^\omega||b$, three peak structures are clearly observed at 0.97, 1.8, and 1.98 THz, while for $E^\omega||c$, a peak at 0.97 THz is prominent. In addition, a broad structure seems to exist at 1.2-1.7 THz for both polarizations in common. A possible origin of the oscillatory structures [Figs. 2(b) and 2(c)] and the narrowband terahertz radiations [Fig. 3(a)] is the impulsive stimulated Raman scattering process, as discussed later. The DCMBI crystal has no inversion symmetry, so that a Raman active mode can be infrared active and then contribute to the terahertz radiation. To demonstrate this mechanism, we performed Raman spectroscopy and IR spectroscopy.

We measured the polarized Raman spectra using *ab*-, *bc*-, and *ac*-surface crystals in the terahertz region, as detailed in the Supplementary Material S4 [16]. According to Raman selection-rule with *mm*2 symmetry, $A_1(z)$ longitudinal-optical (LO) phonon modes become active in $c(bb)\bar{c}$ configuration; Raman spectrum obtained in $c(bb)\bar{c}$ configuration is shown in Fig. 3(b) (circles), in which the five LO modes are identified at 0.99 THz, 1.21 THz, 1.80 THz, 2.03 THz, and 2.40 THz as shown by the vertical arrows. A Raman scattering intensity spectrum of each mode (*i*=1-5) is given by the following formula [18].

$$I(\omega) = \frac{1}{1 - \exp\left(-\frac{\hbar}{k_B T}\right)} \left[1 - \left(\frac{n-1}{n+1}\right)^2\right]^2 \frac{2\hbar\omega_s^4 N^2}{\rho c^4 \omega_v} \sum_i^m \frac{a_i^2 \gamma_i \omega}{(\omega_i^2 - \omega^2) + \gamma_i^2 \omega^2}. \quad (4)$$

Here, $n$ is a refractive index (1.7) at 633 nm [Fig. 1(c)], $\rho$ the density (1.586 kg/m$^3$) [11], $N$ the volume in a primitive cell (4.75×10$^{27}$ m$^{-3}$) [11], $\omega_s$ the center frequency of the scattering light, and $c$ the velocity of light. $\omega_i$, $a_i$, and $\gamma_i$ are the central frequency, Raman polarizability tensor, and damping of the Raman modes, respectively. In Eq. (4), the first term shows the correction with the Bose-Einstein factor, the second term shows the correction for the reflection losses of the excitation light (633 nm) and the scattered light (~633 nm) at the sample surface. The third term dominates the Raman scattering intensity. Using Eq. (4), the Raman spectrum is almost reproduced as shown by the solid line in Fig. 3(b). Spectra of each mode are also shown by the dotted line and color lines in the lower part of Fig. 3(b). In order to evaluate the absolute value of $a_i$, we measured the Raman spectrum around 1330 cm$^{-1}$ of a *c*-axis-oriented diamond for the $c(ab)\bar{c}$ configuration and compared its integrated intensity with that of each Raman band in DCMBI. The $\gamma_i$ and $a_i$ values thus obtained are listed in Table I.

Figure 3(c) shows the spectra of the real ($\varepsilon_1$) and imaginary ($\varepsilon_2$) part of the dielectric constant for $E^\omega||c$ obtained from the terahertz TDS on a *bc* plane in a 397 μm-thick single crystal. We can see sharp peak structures at 0.99, 1.83, and 2.03 THz indicated by vertical arrows in the $\varepsilon_2$ spectrum and the corresponding dispersions in the $\varepsilon_1$ spectrum. Three peak structures at 0.99, 1.83, and 2.03 THz in Fig. 3(c) are also observed in the Raman spectrum in Fig. 3(b), while no IR-active modes corresponding to the Raman peaks at 1.21 THz and 2.40 THz (indicated by black upward arrows) are observed. In order to evaluate the values of the dipole moment $\mu_i$ of each IR-active mode (*i*=1-3), we performed the fitting analysis using three Lorentz oscillators (Supplementary Materials S3 [16]). The calculated $\varepsilon_1$ and $\varepsilon_2$ spectra well reproduce the experimental ones as shown by the solid lines in Fig. 3(c). In this analysis, the high-frequency dielectric constant $\varepsilon_\infty$ is set to be 3.16. Each $\varepsilon_2$ component was also shown in the lower part of Fig. 3(c). The obtained parameters including $\mu_i$ are also listed in Table I.

On the basis of the results presented above, we discuss the terahertz-radiation mechanism in DCMBI. As seen in Figs. 3(a)-(c) and Table I, the central frequencies of the peak structures observed in the terahertz-radiation spectrum for $E^\omega||b$ corresponds with those of Raman and infrared-active phonon modes, which is also detailed in the Supplementary Materials S5 [16]. When the spectral width of the incident femtosecond laser pulse (~10 THz) exceeds the frequency of Raman-active phonons, these modes are coherently excited (Supplementary Materials S6 [16]). This process is called impulsive stimulated Raman scattering (ISRS) [19-21]. If those phonons are also infrared-active, they produce oscillations of corresponding polarizations, resulting in narrowband terahertz radiations as discussed in the study of TeO$_2$ [22]. In the case of the terahertz radiation obtained for $E^\omega||c$, it would be related to the phonon modes with $k$ vector parallel to the $a$-axis, which can be detected by using a (011)-surface crystal, as discussed in the Supplementary Materials S5 [16]. However, obtained crystal has a (100)-surface. Thus, in this work, we focus on the results obtained for $E^\omega||b$.

To demonstrate an ISRS-induced terahertz radiation, we formulate it and evaluate the spectrum of the corresponding $\chi^{(2)}$. The amplitude of the phonon mode is given by $Q(t) = \int_{-\infty}^{t} dt' G_{11}(t-t') \sum_{kl} R_{jk} E_j(t') E_k^*(t')$ [21]. Here, $G_{11}$ is a Green's function and $R_{jk}$ is a Raman tensor. $Q(\omega)$ is expressed by $Q(\omega) = R_{jk} \frac{c^2 k^2 - \varepsilon_\infty \omega^2}{\varepsilon_\infty(\omega^4 + \omega^3(2i\gamma) - \omega^2 \frac{c^2 k^2 - \varepsilon_\infty \omega_{TO}^2}{\varepsilon_\infty} - \omega \frac{2i\gamma c^2 k^2}{\varepsilon_\infty} + \frac{\omega_{TO}^2 c^2 k^2}{\varepsilon_\infty})} E_j(\omega) E_k^*(\omega)$, as detailed in the Supplementary Material S6 [16]. Here, $\gamma$ is a damping constant and $\varepsilon_\infty$ is an infinite dielectric constant. By taking into account the conservation law of the wavevector, i.e., $k$ = 0 and Lyddane-Sachs-Teller relation $\left(\omega_{LO}^2 = \frac{\varepsilon_0 \omega_{TO}^2}{\varepsilon_\infty}\right)$, $Q(\omega)$ is given by $Q(\omega) =$ $-\frac{R_{jk}}{\omega^2-\omega_{\text{LO}}^2+2i\gamma\omega}E_j(\omega)E_k^*(\omega)$ [21]. The nonlinear polarization due to ISRS is represented by $P_i(\omega) = N\mu_i Q(\omega) = \varepsilon_0 \chi_{ijk}^{(2)\text{ISRS}} E_j(\omega) E_k^*(\omega)$, where $e$ is an electrical charge. Here, we assume that only a single excited state with the energy $\omega_\Delta$ contributes to the Raman scattering process and an excitation light is in the transparent region (an off-resonant condition). In this case, $R_{jk}$ can be expressed as $R_{jk} = \frac{e^2}{\varepsilon_0 \hbar^2}\left(\frac{a_{jk}\omega_v}{(\omega_\Delta-\omega_{\text{Laser}})^2+\omega_i^2}\right)$. Then, we get $\chi_{ijk}^{(2)\text{ISRS}}(\omega_i = \omega_j - \omega_k) = -\frac{Ne^3}{\varepsilon_0^2 \hbar^2}\frac{\mu}{\omega^2-\omega_{\text{LO}}^2+2i\gamma\omega}\left(\frac{2a_{jk}\omega_v}{(\omega_\Delta-\omega_{\text{Laser}})^2+\omega_i^2}\right)$. By seting $\omega_\Delta$ and $\omega_{\text{Laser}}$ to be the resonant energy (6.7 eV) derived from the Sellmeier relationship [see Fig. 1(d)] and the laser photon energy (1.55 eV), respectively, and using the values of $\mu, a, \gamma$, and $\omega_{\text{LO}}$ listed in Table I, we calculate the $|\chi_{cbb}^{(2)\text{ISRS}}|$ spectrum, which is shown by the solid line in Fig. 3(d). $|\chi_{cbb}^{(2)\text{ISRS}}|$ is enhanced with resonance to three Raman- and IR-active phonon modes shown by blue arrows and reaches the maxima of ~0.0035 pm/V. The $|\chi_{cbb}^{(2)\text{ISRS}}|$ spectrum is similar to the terahertz-radiation spectrum shown in Fig. 3(a) (blue circles) except for the broad structure at 1.2-1.7 THz. It is reasonable to consider that this broad structure observed for $E^\omega||b$ as well as $E^\omega||c$ originates from the OR, which is a general mechanism of terahertz radiation in noncentrosymmetric media. By using $|\chi_{cbb}^{(2)\text{ISRS}}|$ and taking the detector response function into account [22], we calculate the electric field $E_X(t)$ of the terahertz radiation for $E^\omega||b$, which is shown by the solid line in Fig. 4. The time delay is arbitrary. The calculated curve almost reproduces the experimental data for $E^\omega||b$. This result suggests that the terahertz radiation for $E^\omega||b$ is dominated mainly by the IR-active phonon mode excited viat the ISRS mechanism.

Finally, we discuss the characteristic of ISRS-induced narrowband terahertz radiations in DCMBI, compared to other organic ferroelectrics, in which OR is dominant. In the case of croconic acid, the emitted waveform shows the broad spectrum with the frequency component up to 2 THz. This can be explained by OR [6]. In croconic acids, longitudinal direction of all molecules is parallel to the hydrogen-bonding direction, i.e., ferroelectric polarization. Since the asymmetric π electron configuration is related to the magnitude of $\chi^{(2)}$, this arrangement results in the enhancement of OR. Indeed, $E_{\text{THz}}(0)$ in the croconic acid is comparable to that in ZnTe, which is a typical terahertz emitter. On the other hand, in the case of DCMBI, longitudinal direction of all molecules along the *b*-axis [indicated by blue shaded area in Fig. 1(a)] is perpendicular to the hydrogen-bonding direction, i.e., ferroelectric *c*-axis. This results in the reduction of OR. In addition, in DCMBI, Raman and IR active modes exist in the measured frequency region and thus $E_{\text{THz}}(\omega)$ resonantly enhances due to the ISRS mechanism, as revealed by the present work.

In summary, we successfully observed the emission of the narrowband terahertz waves in an above-room-temperature organic ferroelectric DCMBI upon irradiation of a femtosecond laser pulse. We concluded that the narrowband terahertz-radiation mechanism is due to ISRS and evaluated the ISRS-induced second-order nonlinear optical susceptibility spectrum in the terahertz region.

We thank Mr. M. Tateno for experimental support in an early stage of this study. This work was partly supported by a Grant-in-Aid by MEXT (No. 25247049, 25600072, and 15H03549). M. S. was supported by Japan Society for the Promotion of Science (JSPS) through Program for Leading Graduate Schools (MERIT) and JSPS Research Fellowships for Young Scientists.

**References**

[1] M. E. Lines and A. M. Glass, *Principles and Applications of Ferroelectric and Related Materials* (Clarendon, Oxford, 1977).

[2] Y. R. Shen, *The Principles of Nonlinear Optics* (Wiley, New York, 1984).

[3] S. Horiuchi and Y. Tokura, Organic ferroelectrics, Nat. Mater. **7**, 357-366 (2008).

[4] S. Horiuchi, Y. Tokunaga, G. Giovannetti, S. Picozzi, H. Itoh, R. Shimano, R. Kumai, and Y. Tokura, Above-room-temperature ferroelectricity in a single-component molecular crystal, Nature (London) **463**, 789-792 (2010).

[5] S. Horiuchi, R. Kumai, and Y. Tokura, Hydrogen-bonding molecular chains for high-temperature ferroelectricity, Adv. Mater. **23**, 2098-2103 (2011).

[6] M. Sotome, N. Kida, S. Horiuchi, and H. Okamoto, Visualization of ferroelectric domains in a hydrogen-bonded molecular crystal using emission of terahertz radiation, Appl. Phys. Lett. **105**, 041101 (2014).

[7] W. Guan, N. Kida, M. Sotome, Y. Kinoshita, R. Takeda, A. Inoue, S. Horiuchi, and H. Okamoto, Terahertz radiation by optical rectification in a hydrogen-bonded organic molecular ferroelectric crystal, 2-phenylmalondialdehyde, Jpn. J. Appl. Phys. **53**, 09PD07 (2014).

[8] M. Bass, P. A. Franken, J. F. Ward, and G. Weinreich, Optical rectification, Phys. Rev. Lett. **9**, 446-448 (1962).

[9] M. Tonouchi, Cutting-edge terahertz technology, Nat. Photon. **1**, 97-105 (2007).

[10] A. Schneider, M. Neis, M. Stillhart, B. Ruiz, R. U. A. Khan, and P. Günter, Generation of terahertz pulses through optical rectification in organic DAST crystals: theory and experiment, J. Opt. Soc. Am. B **23**, 1822-1835 (2006).

[11] S. Horiuchi, F. Kagawa, K. Hatahara, K. Kobayashi, R. Kumai, Y. Murakami, and Y. Tokura, Above-room-temperature ferroelectricity and antiferroelectricity in benzimidazoles, Nat. Commun. **3**, 1308 (2012).

[12] M. Sotome, N. Kida, S. Horiuchi, and H. Okamoto, Terahertz radiation imaging of ferroelectric domain topography in room-temperature hydrogen-bonded supramolecular ferroelectrics, ACS Photonics **2**, 1373-1383 (2015).

[13] M. Sotome, N. Kida, Y. Kinoshita, H. Yamakawa, T. Miyamoto, H. Mori, and H. Okamoto, Visualization of a nonlinear conducting path in an organic molecular ferroelectric by using emission of terahertz radiation, Phys. Rev. B **95**, 241102(R) (2017).

[14] Y. Kinoshita, N. Kida, M. Sotome, R. Takeda, N. Abe, M. Saito, T. Arima, and H. Okamoto, Visualization of ferroelectric domains in boracite using emission of terahertz radiation, Jpn. J. Appl. Phys. **53**, 09PD08 (2014).

[15] Y. Kinoshita, N. Kida, M. Sotome, T. Miyamoto, Y. Iguchi, Y. Onose, and H. Okamoto, Terahertz radiation by subpicosecond magnetization modulation in the ferrimagnet $LiFe_5O_8$, ACS Photonics **3**, 1170-1175 (2016).

[16] See, supplementary information for ferroelectric domain structure visualized by terahertz radiation imaging, second-order nonlinear optical susceptibility, estimation of coherence length, polarized Raman scattering spectra, Comparison terahertz emission spectra with Raman and dielectric constant spectra, and impulsive stimulated Raman scattering process.

[17] R. Takeda, N. Kida, M. Sotome, Y. Matsui, and H. Okamoto, Circularly polarized terahertz radiation from a eulytite oxide by a pair of femtosecond laser pulses, Phys. Rev. A **89**, 033832 (2014).

[18] M. Grimsditch and M. Cardona, Absolute cross-section for Raman scattering by phonons in silicon, Phys. Status Solidi B **102**, 155-161 (1980).

[19] Y.-X. Yan and K. A. Nelson, Impulsive stimulated light scattering. I. General theory, J. Chem. Phys. **87**, 6240-6256 (1987).

[20] Y.-X. Yan and K. A. Nelson, Impulsive stimulated light scattering. II. Comparison to frequency-domain light-scattering spectroscopy, J. Chem. Phys. **87**, 6257-6265 (1987).

[21] T. P. Dougherty, G. P. Wiederrecht, and K. A. Nelson, Impulsive stimulated Raman scattering experiments in the polariton regime, J. Opt. Soc. Am. B **9**, 2179-2189 (1992).

[22] M. Sotome, N. Kida, R. Takeda, and H. Okamoto, Terahertz radiation induced by coherent phonon generation via impulsive stimulated Raman scattering in paratellurite, Phys. Rev. A **90**, 033842 (2014).

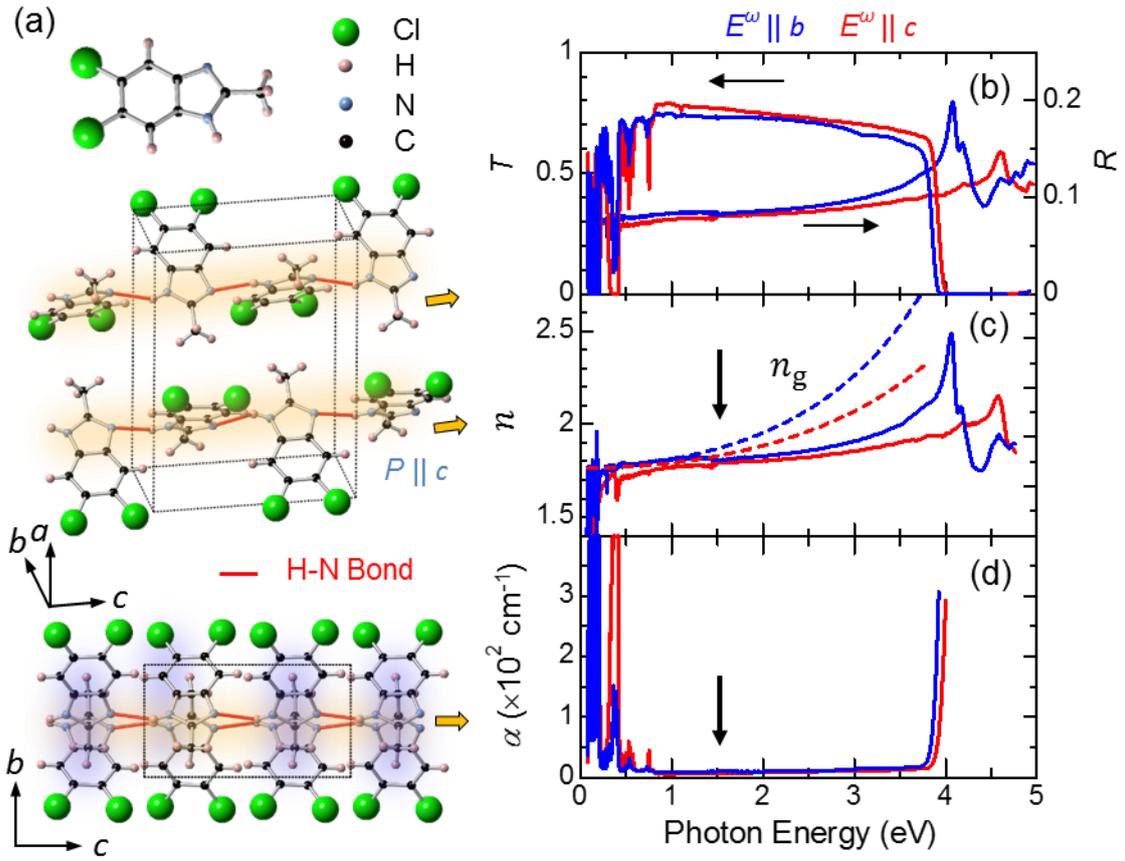

FIG. 1. (a) Schematics of molecular and crystal structures of 5,6-dichloro-2-methylbenzimidazole. Spontaneous polarization $P$ (arrows) appears along the H-N bond direction ($c$-axis) indicated by the red line. The imidazole molecules are indicated by the shaded blue area. (b) Polarized transmittance $T$ and reflectivity $R$, (c) refractive index $n$, and (d) absorption coefficient $\alpha$ spectra for $E^{\omega} \parallel b$ (blue lines) and $E^{\omega} \parallel c$ (red lines). The group refractive index $n_g$ spectra are also shown by dotted lines in (c). The vertical arrows indicate the photon energy of a femtosecond laser pulse used here.

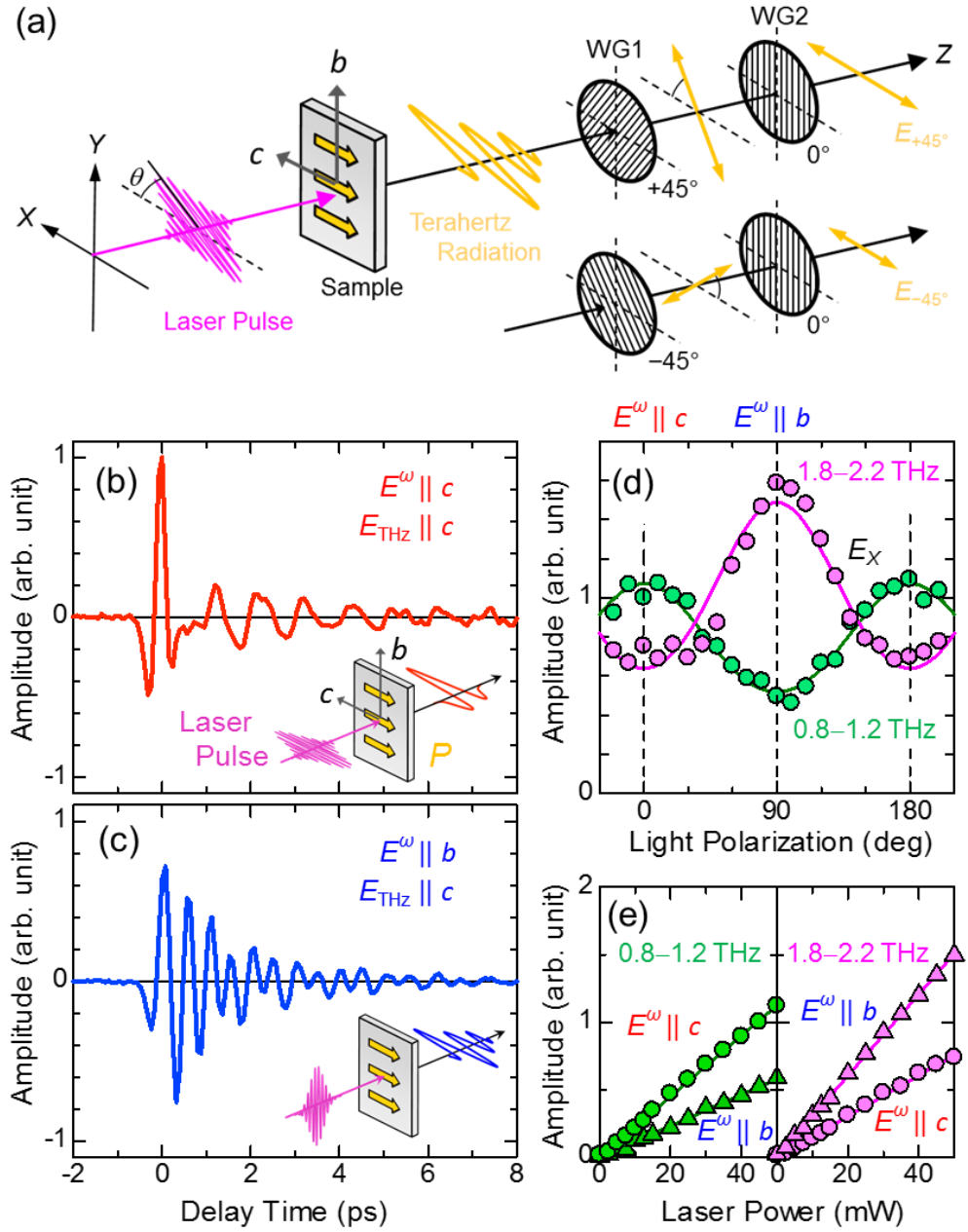

FIG. 2. (a) Schematic for vector analysis. $X$-axis component of terahertz waveforms for (b) $E^{\omega}\|c$ and (c) $E^{\omega}\|b$. (d) Light-polarization angle $\theta$ and (e) laser power dependences of the terahertz electric-fields in the frequency range of 0.8-1.2 THz (green) and 1.8-2.2 THz (pink).

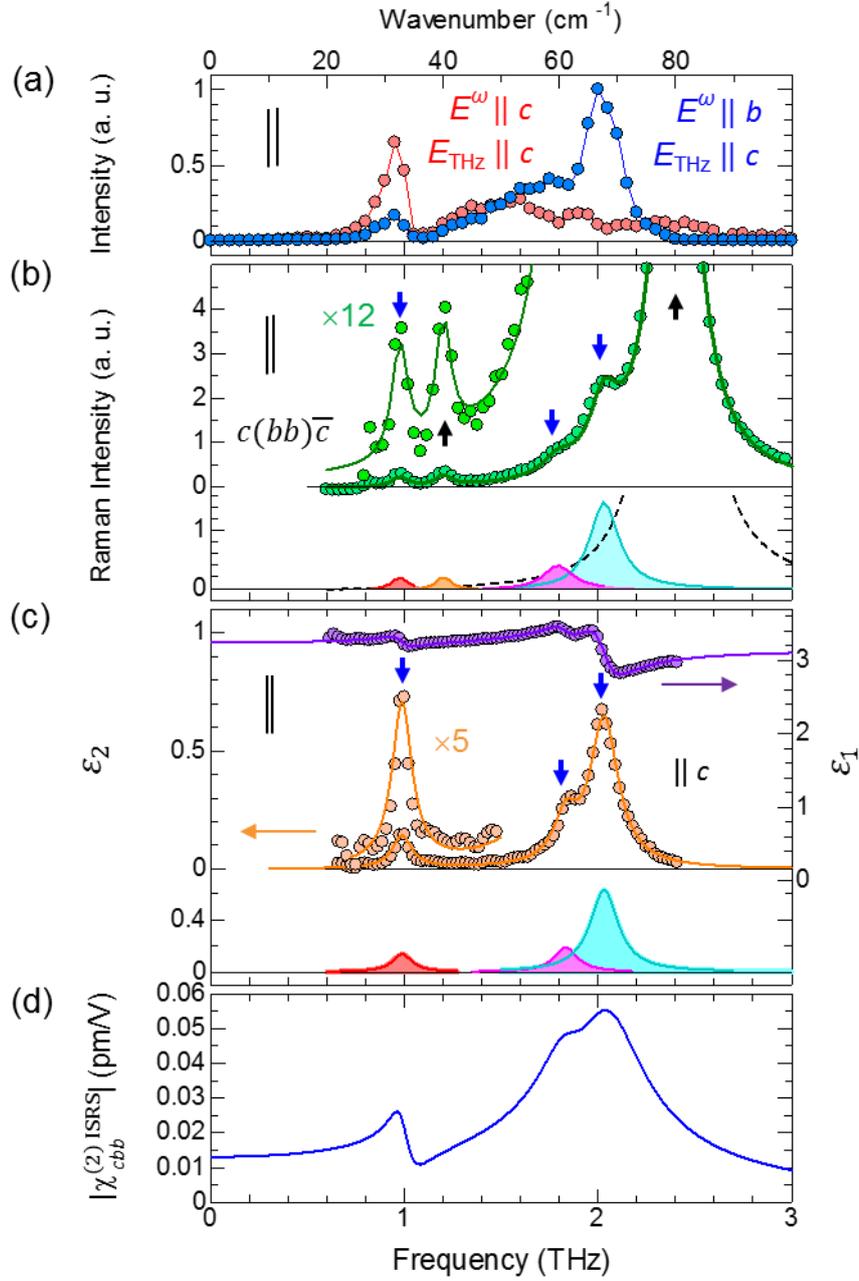

FIG. 3. (a) Intensity spectra of terahertz radiation for $E^{\omega}\|c$ and $E^{\omega}\|b$. (b) Raman scattering spectrum in $c(bb)\bar{c}$ configuration, which can be reproduced by assuming five Raman modes (shade area and dotted line). (c) Real $\varepsilon_1$ and imaginary $\varepsilon_2$ parts of the dielectric constant spectrum along the $c$-axis, which can be reproduced by three Lorentz oscillators (shade area). (d) Evaluated spectrum of the second-order nonlinear susceptibility $|\chi^{(2)\text{ISRS}}|$ due to the impulsive stimulated Raman scattering.

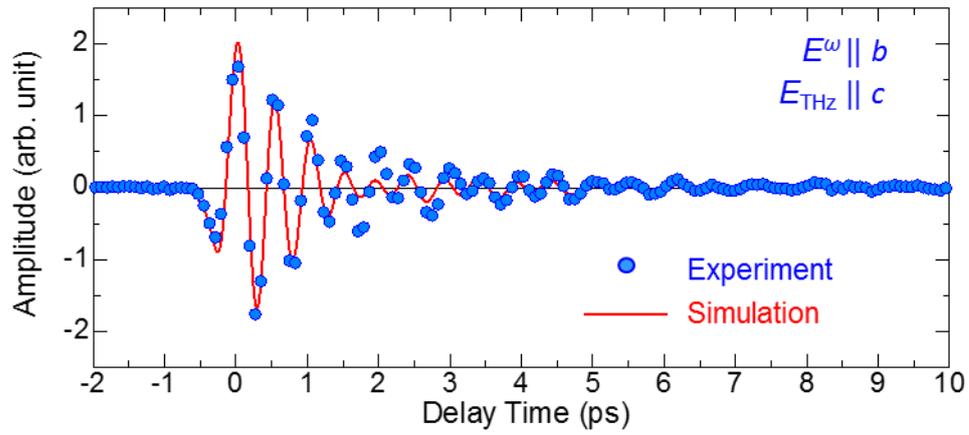

FIG. 4. Terahertz waveform obtained for $E^\omega \| b$ (circles) and simulation curve originating from the impulsive stimulated Raman scattering (solid line).

Table I. Obtained fitting parameters of Raman-active and infrared-active phonon modes and comparison with terahertz radiation polarized along the *c*-axis for $E^{\omega}||b$.

| Terahertz | Raman | | | Infrared | | |
|---|---|---|---|---|---|---|
| $E^{\omega}||b$, $E_{\text{THz}}||c$ | $c(bb)\bar{c}$ | | | $||c$ | | |
| $\omega/2\pi$ (THz) | $\omega/2\pi$ (THz) | $\gamma/2\pi$ (THz) | $a$ (Å$^2$) | $\omega/2\pi$ (THz) | $\gamma/2\pi$ (THz) | $\mu$ (Å$^2$) |
| 0.97±0.05 | 0.99 | 0.0981 | 0.034 | 0.99 | 0.126 | 0.004 |
| – | 1.21 | 0.090 | 0.038 | – | – | – |
| 1.80±0.05 | 1.80 | 0.189 | 0.092 | 1.83 | 0.148 | 0.007 |
| 1.98±0.05 | 2.03 | 0.180 | 0.180 | 2.03 | 0.177 | 0.166 |
| – | 2.40 | 0.203 | 0.677 | – | – | – |

**Supplemental Material**

**Narrowband terahertz radiation by an impulsive stimulated Raman scattering in an above-room-temperature organic ferroelectric benzimidazole**

M. Sotome[1], N. Kida[1,*], S. Horiuchi[2], and H. Okamoto[1,3]

[1]*Department of Advanced Materials Science, The University of Tokyo, 5-1-5 Kashiwa-no-ha, Chiba 277-8561, Japan*

[2]*National Institute of Advanced Industrial Science and Technology (AIST), Tsukuba 305-8562, Japan*

[3]*AIST-U. Tokyo Advanced Operando-Measurement Technology Open Innovation Laboratory, National Institute of Advanced Industrial Science and Technology, Chiba 277-8568, Japan*

* kida@k.u-tokyo.ac.jp

**Content**

**S1. Ferroelectric domain structure visualized by terahertz radiation imaging**

**S2. Second-order nonlinear optical effect**

**S3. Estimation of coherence length**

**S4. Polarized Raman scattering spectra**

**S5. Comparison terahertz emission spectra with Raman and dielectric constant spectra**

**S6. Impulsive stimulated Raman scattering (ISRS) process**

**S1. Ferroelectric domain structure visualized by terahertz radiation imaging**

Ferroelectric domain structure is usually determined by the symmetry of the crystal [1]. However, it is actually affected by surface roughness, sample size, and growth conditions of the crystals. Indeed, complicated domain structures separated by a charged head-to-head (tail-to-tail) domain wall (DW), which is energetically unstable compared to a 180° DW, were clearly discerned in the virgin state of as-grown organic ferroelectric crystals [2, 3]. In this context, it is indispensable to detect the ferroelectric domain structure of DCMBI in the virgin state.

In order to map out ferroelectric domain structures over a DCMBI sample, we used terahertz radiation imaging method [2-6]. We identified that the phase of the terahertz electric field radiated from a ferroelectric crystal depends on the polarization direction. Thus, we can visualize the ferroelectric domains and DWs by measuring the position dependence of the electric field at 0 ps $[E_{\text{THz}}(0)]$ in the time-domain waveform of terahertz radiations. In this experiment, the electric field of the incident femtosecond laser pulse was set parallel to the $c$-axis. We detected $E_{\text{THz}}(0)$ and obtained the ferroelectric domain image at all the area of the sample by a raster scan. We used four samples with a largest $bc$-surface, the size and thickness of four samples are described in Fig. S1. The terahertz radiation images of four samples are shown in lower panels of Figs. S1(a) to S1(d) together with the optical images [upper panels of Figs. S1(a) to S1(d)]. The flat surface region of the crystal is surrounded by a dotted line in both figures. The color bar shown in the right-hand side of the figure indicates the amplitude and sign of $E_{\text{THz}}(0)$, which correspond to the magnitude and direction of the ferroelectric polarization, respectively. In terahertz radiation images obtained in two larger-size samples [lower panels of Figs. S1(a) and S1(b)], we observe a nearly single-domain state over a sample (~3000 μm × ~600 μm) with high homogeneity along the $c$-axis (ferroelectric direction), which is separated by a 180° DW (indicated by the white region). On the other hand, in other smaller-size samples [lower panels of Figs. S1(c) and S1(d)], we observe distinct domain structures, which are separated by charged tail-to-tail (head-to-head) DWs. Especially, there is a domain boundary, which can be seen as a white crossed line in the terahertz radiation image shown in Fig. S1(c). The inclination of these lines is ~±30° from the *c*-axis. These results suggest that the domain pattern depends on the crystal-volume or the crystal-length along the *c*-axis. The single-domain state separated by a stable 180°

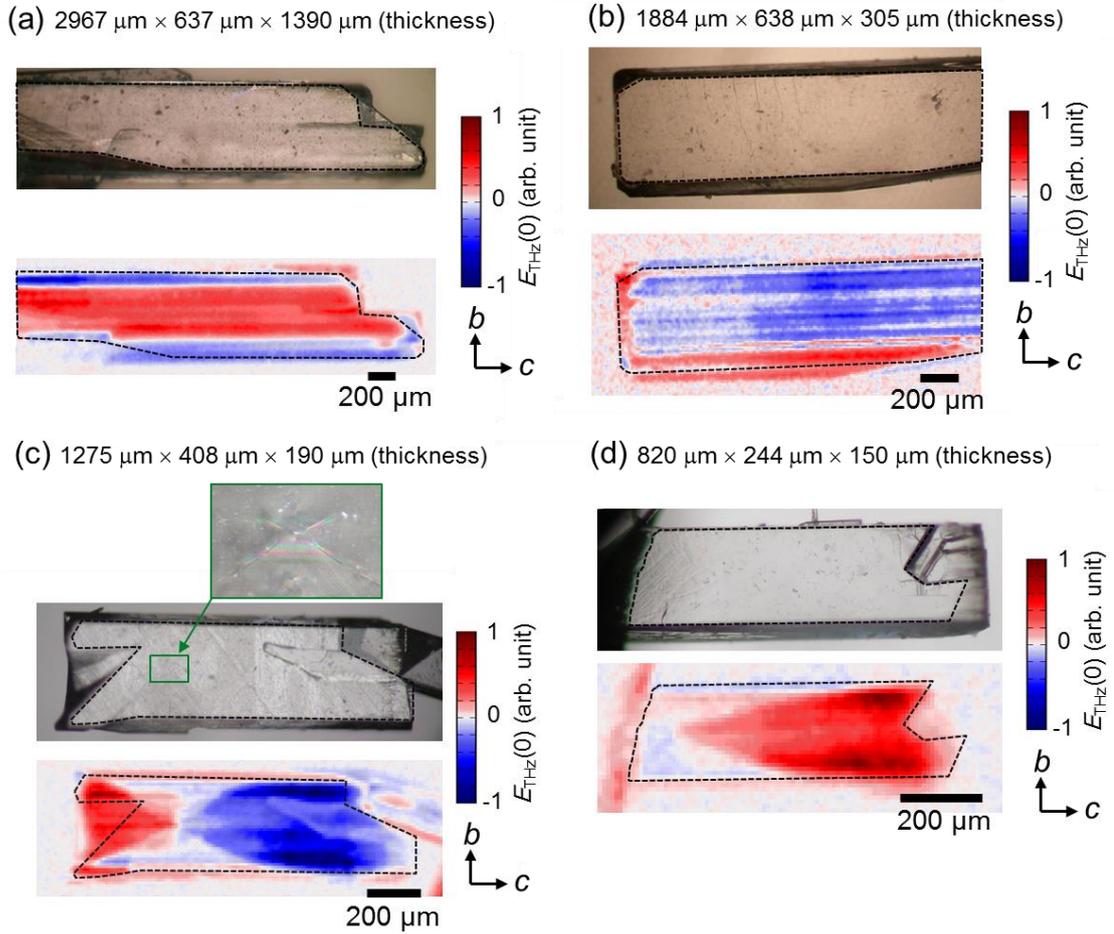

FIG. S1. Optical image (upper panel) and terahertz radiation image (lower panel) obtained in four crystals with the largest *bc*-surface. The size and the thickness are indicated in (a) to (d). The flat surface region of the crystal is surrounded by a dotted line. The crossed-Nicol image of the region indicated by the green box is shown in (c).

DW is realized in two samples with a relatively large crystal-length along the $c$-axis (~2000 μm). On the other hand, unstable charged DWs show up in other samples with a relatively small crystal-length along the $c$-axis (~1000 μm). Such a correlation between crystal-size (or volume) and domain pattern would be related to the orientation of DW during the crystal growth. Namely, the domain boundary as observed in the terahertz radiation image would be a starting point for crystal growth. In order to confirm this assumption, we measured the domain boundary region indicated by the green box by using the optical microscope in the crossed-Nicol geometry. The obtained image is shown in the upper region of Fig. S1(c). From this image, one can be discerned a clear crossed line, which is inclined by ~±30° from the $c$-axis. This is a nearly same to the crossed line observed in the terahertz radiation image. Thus, ferroelectric pattern observed in DCMBI is determined by the kinetics of the crystal growth.

In the terahertz radiation experiments presented in main text, a femtosecond laser pulse with a spot diameter of ~25 μm is irradiated within a single-domain state. In addition, it is noticed that the coherence length, which is discussed lately in the Supplementary Material S3, exceeds or is comparable to the thickness of four samples used in terahertz radiation imaging experiments.

**S2. Second-order nonlinear optical effect**

Since DCMBI is transparent to the excitation light used in terahertz radiation experiments (1.55 eV) [see Fig. 1(d)], the observed terahertz radiation can be ascribed to the second-order optical nonlinearity characterized by the second-order nonlinear susceptibility $\chi^{(2)}$. In the transparent media, the polarization $P^{(2)}$ induced by an irradiation with a laser pulse below bandgap is represented by

$$P_i^{(2)} = \epsilon_0 \chi_{ijk}^{(2)} E_j^\omega E_k^{\omega *}, \quad (S1)$$

where $\epsilon_0$ is the vacuum permittivity and $E^\omega$ is the light electric-field [7]. Since a femtosecond laser pulse has a finite spectral width ~10 THz, the mixing of different frequency components induces the corresponding $P^{(2)}$, resulting in the emission of broadband terahertz waves [8].

The space group of DCMBI is *mm*2, thus non-zero components of $\chi^{(2)}$ are $\chi_{aac}, \chi_{aca}, \chi_{bbc}, \chi_{bcb}, \chi_{caa}, \chi_{cbb}$, and $\chi_{ccc}$, where *a*, *b*, and *c* stand for crystallographic axes [7]. Then, $P^{(2)}$ using the contradicted *d* tensor is expressed as

$$P^{(2)} = \begin{pmatrix} P_a^{(2)} \\ P_b^{(2)} \\ P_c^{(2)} \end{pmatrix} = \epsilon_0 \begin{pmatrix} 2d_{31} E_a^\omega E_c^{\omega *} \\ 2d_{32} E_b^\omega E_c^{\omega *} \\ d_{31} E_a^\omega E_a^{\omega *} + d_{32} E_b^\omega E_b^{\omega *} + d_{33} E_c^\omega E_c^{\omega *} \end{pmatrix}. \quad (S2)$$

When the *c*-axis of the crystal with a largest *bc*-surface is set parallel to the *X*-axis (horizontal-axis) in laboratory coordinate [Fig. 2(a)], *X*-axis electric-field component $E_X(\omega)$ of the terahertz radiation is given by

$$E_X(\omega) \propto (d_{32} \sin^2\theta + d_{33} \cos^2\theta) I, \quad (S3)$$

where $I$ is the laser intensity and $\theta$ is the angle of the electric field of the incident laser pulse measured from the *X*-axis.

To investigate the angle ($\theta$) dependence of the terahertz radiation, we rotate the direction of the electric field ($E^\omega$) of the incident femtosecond laser pulse by using a half-wave plate and evaluate the *X*- and *Y*-axes components [$E_X(t)$ and $E_Y(t)$] of terahertz electric-fields $E_{\text{THz}}(t)$ by means of the vector analysis using two wire-grid polarizers (WG1 and WG2), as schematically shown in Fig. 2(a). With respect to the *X*-axis, the angle of WG2 is set 0°, while that of WG1 is set +45° or −45°. We measure the corresponding electric fields, $E_{+45°}(t)$ and $E_{-45°}(t)$, and obtain $E_X(t)$ and $E_Y(t)$ from the relations, $E_X(t) = \frac{1}{\sqrt{2}}(E_{+45°}(t) - E_{-45°}(t))$ and $E_Y(t) = \frac{1}{\sqrt{2}}(E_{+45°}(t) + E_{-45°}(t))$, respectively. The further information concerning the calculation procedure of $E_X(t)$ and $E_Y(t)$ by the vector analysis is reported in Refs. [4, 6]. Figure 2(d) shows $\theta$ dependence of $E_X(\omega)$ integrating in the frequency range of 0.8-1.2 THz and 1.8-2.2 THz, which are indicated by green and pink circles, respectively. Equation (S3) well reproduces $\theta$ dependence of $E_X(\omega)$, which are indicated by solid lines. Combined with their $I$ dependences shown in Fig. 2(e) for $E^\omega||c$ and $E^\omega||b$, we clearly demonstrate that second-order nonlinear optical effect dominates for the light-induced terahertz radiation from DCMBI.

## S3. Estimation of coherence length

In order to estimate the coherence length $(l_c)$ for terahertz radiation, we measured the polarized transmission spectra in the terahertz region by terahertz time-domain spectroscopy and estimated the complex optical constants. Our estimation procedure is reported in Ref. [9]. In this experiment, we used a 397 μm-thick crystal with a largest *bc*-surface. Figure S2(a) shows the spectra of the refractive index $n_{THz}$ along the *c*- and *b*-axes, which are shown by orange and purple circles, respectively. In both configurations, the clear dispersion structures are discerned. Accordingly, the peak structures appear in spectra of the imaginary part of the dielectric constant $\varepsilon_2$ at 0.99 THz, 1.84 THz, and 2.02 THz for the *c*-axis, while at 0.86 THz, 1.25 THz, and 1.8 THz for the *b*-axis, which are shown by orange and purple circles, respectively, in Fig. S2(b). In order to estimate the dipole moment $\mu_i$ of each IR-active mode (*i*=1-3), we performed the fitting analysis using three Lorentz oscillators, which are given by

$$\epsilon_1(\omega) = \epsilon_\infty + \sum_{i=1}^{3} \left[ \frac{f_i \omega_i^2 (\omega_i^2 - \omega^2)}{(\omega_i^2 - \omega^2)^2 + \gamma_i^2 \omega^2} \right], \qquad (S4)$$

$$\epsilon_2(\omega) = \sum_{i=1}^{3} \left[ \frac{f_i \omega_i^2 \gamma_i \omega}{(\omega_i^2 - \omega^2)^2 + \gamma_i^2 \omega^2} \right], \qquad (S5)$$

with $f_i$, which is represented by

$$\mu_i = \frac{\epsilon_0 \hbar \omega_i}{2Ne^2} f_i. \qquad (S6)$$

The obtained $\mu_i$ are listed in Table I. We also indicate the value of the group refractive index $n_g$ at 1.55 eV by horizontal lines in Fig. S2(a); 1.88 ($E^\omega||b$) and 1.83 ($E^\omega||c$), which are obtained by Sellmeier relation from the refractive index spectra shown in Fig. 1(c).

$l_c$ is represented by $\frac{\lambda}{2} \frac{1}{|n_g - n_{THz}|}$, where $\lambda$ is the wavelength of the emitted terahertz waves. We estimated the spectra of $l_c$ along the $c$- and $b$-axes, which are shown by orange and purple circles in Fig. S2(c), respectively. In both configurations, the increase of $l_c$ in the narrow-frequency region occurs by satisfying the phase-matching condition ($n_g \sim n_{THz}$) due to the presence of the sharp dispersion of the phonon modes in the terahertz region, as seen in Fig. S2(a). This would produce the narrowband terahertz radiation, which is indeed observed near the phonon-polariton resonance in ZnTe [10] and the $F_2$ mode in Bi$_4$Ge$_3$O$_{12}$ [11]. In our terahertz radiation experiments for $E^\omega||c$ and $E^\omega||b$, we detected the power spectra of the terahertz radiations polarized along the $c$-axis. Thus, the observed terahertz radiation may be related to the increase of $l_c$ along the $c$-axis; i.e., the peak structures in the spectrum of $l_c$ may appear in the power spectra of terahertz radiations. However, this possibility is clearly excluded. The sample thickness used in terahertz radiation experiments was set to be 150 μm [horizontal red line in Fig. S2(c)]. This value is shorter than the minimum value of $l_c$, i.e., 1 mm at 2.1

THz. Thus, in our experimental condition, the good phase-matching condition is satisfied in the measured frequency region (0–3 THz). In addition, the observed anisotropic power spectra of terahertz radiation obtained for $E^{\omega}||c$ and $E^{\omega}||b$ [Fig. 3(a)] cannot be reproduced by the spectrum of $l_c$ along the $c$-axis. These results clearly rule out the possibility of the phase-matching-induced narrowband terahertz radiation in DCMBI.

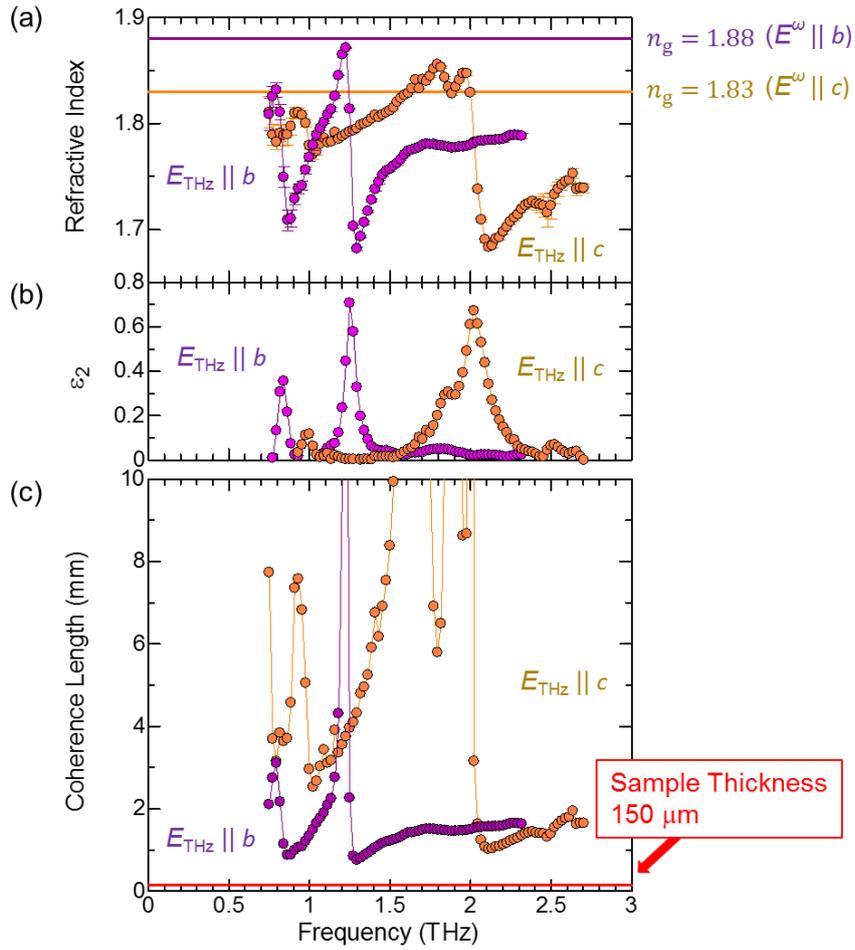

FIG. S2. Spectra of (a) Refractive index and (b) imaginary part of the dielectric constant $\varepsilon_2$ along the $b$- and $c$-axes. (b) Calculated coherence length along the $b$- and $c$-axes, which exceeds the sample thickness (150 μm indicated by the horizontal red line) used in the terahertz radiation experiments.

**S4. Polarized Raman scattering spectra**

Raman tensors of DCMBI with the point group of *mm*2 are given by

$$A_1(z) = \begin{pmatrix} a & 0 & 0 \\ 0 & b & 0 \\ 0 & 0 & c \end{pmatrix}, \qquad A_2 = \begin{pmatrix} 0 & d & 0 \\ d & 0 & 0 \\ 0 & 0 & 0 \end{pmatrix},$$

$$B_1(x) = \begin{pmatrix} 0 & 0 & e \\ 0 & 0 & 0 \\ e & 0 & 0 \end{pmatrix}, \qquad B_2(y) = \begin{pmatrix} 0 & 0 & 0 \\ 0 & 0 & f \\ 0 & f & 0 \end{pmatrix}, \qquad (S7)$$

where *a*, *b*, *c*, *d*, *e*, and *f* are the constant. We listed in Table SI the Raman selection-rule given by Eq. (S7). In order to assign the Raman modes of DCMBI in the terahertz region, we measured the polarized Raman spectra using *bc*-, *ab*-, and *ac*-surface crystals in the backscattering geometry, the results of which in the frequency region of 0.5–3 THz are summarized in Figs. S3(a), S3(b), and S3(c), respectively. According to Raman selection-rule listed in Table S1, $B_1(x)$ mode is active for $b(ac)\bar{b}$ and $b(ca)\bar{b}$ configurations. The obtained Raman spectra in $b(ac)\bar{b}$ and $b(ca)\bar{b}$ configurations are shown by pink and black lines in Fig. S3(c), respectively. There are peak structures at 0.82 THz, 1.3 THz, 2.4 THz, and 2.67 THz in both spectra, which are all assigned to this mode. On the other hand, $B_2(y)$ mode is active for $a(bc)\bar{a}$ and $a(cb)\bar{a}$ configurations, the results of which are shown by orange and yellow lines in Fig. S3(a), respectively. We observe the peak structures at 0.82 THz, 1.3 THz, 2.3 THz, and 2.67 THz in both spectra and assigned them to this mode. In Raman spectra obtained for $c(ab)\bar{c}$ and $c(ba)\bar{c}$ configurations, which are shown by sky blue and brown lines in Fig. S3(b), only $A_2$ mode appears. $A_2$ modes can be discerned at 0.82 THz, 1.2 THz, 1.77 THz, and 2.4 THz. $A_1(z)$ transverse-optical (TO) phonon mode becomes active for $a(bb)\bar{a}, a(cc)\bar{a}, b(aa)\bar{b},$ and $b(cc)\bar{b}$ configurations. As seen in the Raman spectrum in $c(bb)\bar{c}$ configuration, which is shown by a red line in Fig. S3(a), we observe two clear peak structures at 0.99 THz and 2.4 THz and tiny peak structure at 1.77 THz indicated by an asterisk. The former two peak structures are also discerned in the Raman spectrum in $b(cc)\bar{b}$ configuration, which is shown by a purple line in Fig. S3(c). Thus, the peak structures at 0.99 THz and 2.4 THz are assigned to the $A_1(z)$ TO phonon modes. The peak structure (1.77 THz) is due to the leakage of the $A_2$ mode, which is seen in the Raman spectra in $c(ab)\bar{c}$ and $c(ba)\bar{c}$ configurations [Fig. S3(b)]. In the Raman spectrum in $b(cc)\bar{b}$ configuration [Fig. S3(c)], there is a tiny peak structure at 0.82 THz, which is assigned to the leakage of the $B_1$ mode, as discerned in the Raman spectra in $b(ac)\bar{b}$ and $b(ca)\bar{b}$ configurations. On the other hand, $A_1(z)$ longitudinal-optical (LO) phonon mode becomes active for $c(aa)\bar{c}$ and $c(bb)\bar{c}$ configurations, the obtained Raman spectra of which are shown by blue and green lines in Fig. S3(b), respectively. In both configurations, peak structures show up at 0.99 THz, 1.21 THz, 1.80 THz, 2.06 THz, and 2.38 THz, which are assigned to $A_1(z)$ LO modes. A tiny peak structure at 0.82 THz obtained for $c(bb)\bar{c}$ configuration is the leakage of $A_1$ mode, which is indicated by an asterisk. In $a(bb)\bar{a}$ and $b(aa)\bar{b}$ configurations, where $A_1(z)$ TO mode is active, tiny peak structures at 1.2 THz and 2.0 THz indicated by asterisks are observed as shown by a pink line in Fig. S3(a) and by a gray line in Fig. S3(c), respectively. These structures are due to leakages of $A_1(z)$ LO modes. We listed in Table SI results of the Raman mode assignments.

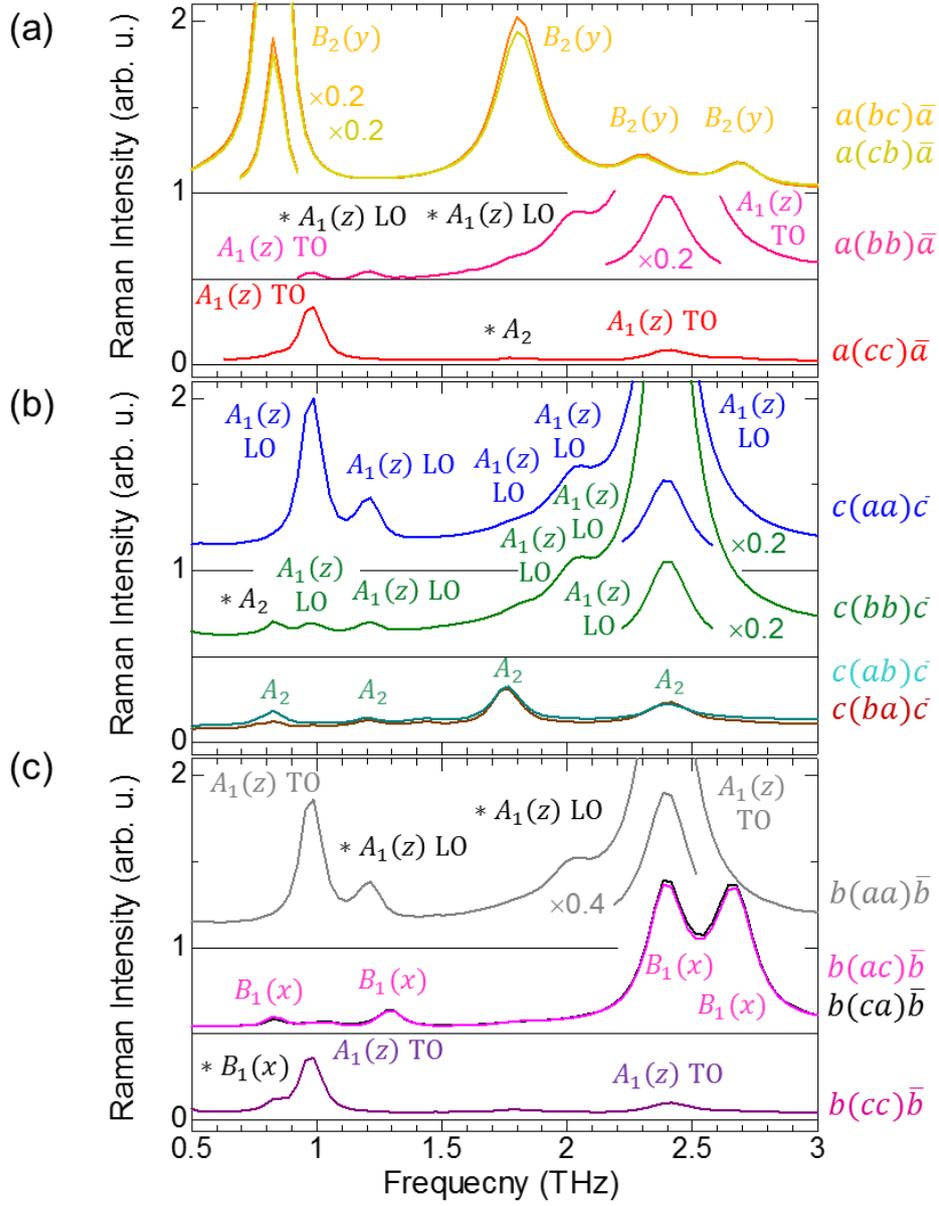

FIG. S3. Polarized Raman scattering spectra obtained using (a) *bc*-, (b) *ab*-, and (c) *ac*-surface crystals in the backscattering geometry.

## S5. Comparison terahertz emission spectra with Raman and dielectric constant spectra

Figure S4(a) shows the intensity power spectra of the emission of the terahertz wave for the electric field $E^\omega$ of the incident laser pulse parallel to the $c$- and $b$-axes ($E^\omega||c$ and $E^\omega||b$), respectively, which are also presented in Fig. 3(a). We observe three peak structures at 0.97 THz, 1.8 THz, and 1.98 THz for $E^\omega||b$, while a peak at 0.97 THz is prominent for $E^\omega||c$. For comparison, we measured the polarized optical spectra in transmission geometry by using terahertz time-domain spectroscopy. The obtained $\varepsilon_2$ spectra along the $b$- and $c$-axes, which are indicated by purple and orange lines, respectively, are shown in Fig. S4(b) [$\varepsilon_2$ spectra are identical to those shown in Fig. 3(c) and Fig. S3(b)]. We observed peak structures at 0.86 THz, 1.25 THz, and 1.8 THz in the $\varepsilon_2$ spectrum along the *b*-axis. On the other hand, peak structures at 0.99 THz, 1.84 THz, and 2.02 THz are observed in the $\varepsilon_2$ spectrum along the *c*-axis. We show in Fig. S4(c) the polarized Raman scattering spectra in $c(bb)\bar{c}$ and $a(cc)\bar{a}$ configurations, in which $A_1(z)$ LO and TO modes are active, respectively, both of which are identical to those shown in Fig. S3. $A_1(z)$ LO modes appear at 0.99 THz, 1.21 THz, 1.80 THz, 2.06 THz, and 2.38 THz, while $A_1(z)$ TO modes at 0.99 THz and 2.4 THz.

It should be noticed that the peak positions of two $A_1(z)$ LO modes in the Raman spectrum at 0.99 THz and 2.06 THz nearly coincide with those of the infrared-active phonon modes discerned in the $\varepsilon_2$ spectrum along the *c*-axis at 0.99 THz and 2.02 THz. Furthermore, the central frequencies of the three peak structures observed in the terahertz-radiation spectrum for $E^\omega||b$ [at 0.97 THz and 1.98 THz] nearly corresponds with those of Raman and infrared-active LO phonon modes. This clearly indicates that both Raman- and infrared-active modes contribute to the terahertz radiation.

In the backscattering geometry using an *ab*-surface crystal, LO modes in $A_1(z)$ symmetry can be detected for $c(bb)\bar{c}$ and $c(aa)\bar{c}$ configurations [Fig. S2(b) and Table SI]. These LO modes with wavevector (*k*) vector parallel to the *c*-axis, are related to the emission of the terahertz radiation obtained for $E^\omega||b$ [blue circles in Fig. S4(a)], as discussed above. On the other hand, the emission of the terahertz radiation obtained for $E^\omega||c$ [pink circles in Fig. S4(a)] would be related to the LO modes with *k* vector parallel to the *a*-axis. In this case, LO modes can be detected using a (011)-surface crystal for $a(xx)\bar{a}$ and $a(xy)\bar{a}$ configurations, where $x$ and $y$ are parallel to $[01\bar{1}]$ and $[011]$ of the crystal, respectively. However, it is not easy to obtain a (011)-surface crystal. Thus, in this work, we focus on the results obtained for $E^\omega||b$.

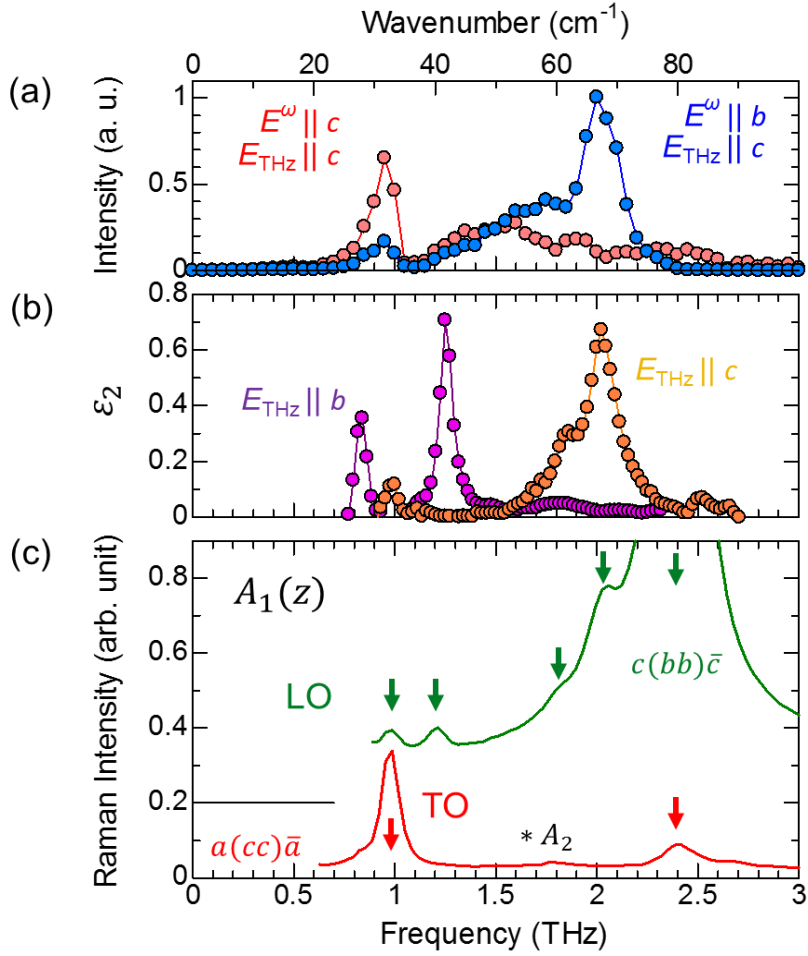

FIG. S4. (a) Intensity power spectra of the emission of the terahertz wave polarized along the $c$-axis for $E^{\omega}||b$ and $E^{\omega}||c$. (b) The obtained spectra of the imaginary part of the dielectric constant $\varepsilon_2$ along the $b$- and $c$-axes. (c) The polarized Raman scattering spectra in $c(bb)\bar{c}$ and $a(cc)\bar{a}$ configurations, in which $A_1(z)$ LO and TO modes becomes active, respectively.

## S6. Impulsive stimulated Raman scattering (ISRS) process

Here, we survey the dispersion relation of the polar optical modes in solids, which is originally discussed in Ref. [12], and also describe its application to understand the impulsive stimulated Raman scattering (ISRS) process developed in Ref. [13]. First, we consider the dispersion of the diatomic cubic crystal with one polar mode [12] and derive

Huang's equations [14]. The equations of motion with the coupling constant $k$ with respect to anions (−) and cations (+) are given by

$$m_- \ddot{u}_- = 2k(u_+ - u_-) - eE_{Loc}, \quad (S8)$$

$$m_+ \ddot{u}_+ = 2k(u_- - u_+) + eE_{Loc}, \quad (S9)$$

respectively. Here, $m_-$ and $m_+$ are masses of anions and cations, respectively. $u_-$ and $u_+$ stand for the lattice displacement coordinates of anions and cations, respectively. $e$ is the effective charge and $E_{Loc}$ is the local field, which is given by $E_{Loc} = E + 4\pi/3 P$, where $E$ is the electric field and $P$ is the polarization. By using effective mass $M = m_- m_+/(m_- + m_+)$, Eqs. (S8) and (S9) become

$$M(\ddot{u}_+ - \ddot{u}_-) = -2k(u_+ - u_-) + eE_{Loc}, \quad (S10)$$

$P$ of the unit cell volume $V$ consists of the electric-dipole originating from the displacement $(u_+ - u_-)$ and $E_{Loc}$, which is given by

$$PV = e(u_+ - u_-) + E_{Loc}(\alpha_+ - \alpha_-), \quad (S11)$$

where $\alpha_-$ and $\alpha_+$ are the polarizabilities of anions and cations, respectively. By using general displacement coordinate $Q = \sqrt{M/V}(u_+ - u_-)$ and $Q$, $P$, and $E$ are all proportional to $\exp(-i\omega t)$, Eqs. (S10) and (S11) simply become

$$-\omega^2 Q = B^{11} Q + B^{12} E, \quad (S12)$$

$$P = B^{21} Q + B^{22} E, \quad (S13)$$

where $B^{11}, B^{12}, B^{21}$, and $B^{22}$ are coefficients. For the transverse optic (TO) mode, where $P$ is perpendicular to the light $k$ vector, $E$ is given by

$$E = \frac{4\pi}{n^2 - 1} P, \quad (S14)$$

where $n$ is the refractive index. For the longitudinal optic (LO) mode, where $P$ is parallel to the light $k$ vector, $E$ is given by

$$E = 4\pi P. \quad (S15)$$

Equations (S12)-(S15) are so-called Huang's equations [13], which satisfy the following condition in the case of TO modes;

$$\begin{vmatrix} -\omega^2 - B^{11} & -B^{12}\dfrac{4\pi}{n^2-1} \\ -B^{21} & -\dfrac{4\pi B^{22} - n^2 + 1}{n^2-1} \end{vmatrix} = 0. \qquad (S16)$$

Then, we obtain the following relation;

$$n^2 = 1 + 4\pi B^{22} - \dfrac{4\pi B^{12} B^{22}}{-\omega^2 - B^{11}}. \qquad (S17)$$

By comparing Eq. (S17) with the conventional expression of $n^2 = \varepsilon_\infty + \omega_{TO}^2(\varepsilon_0 - \varepsilon_\infty)/(\omega_{TO}^2 - \omega^2)$, $B^{11}$ and $B^{22}$ are given by $-\omega_{TO}^2$ and $(\varepsilon_\infty - 1)/4\pi$, respectively.

Taking into account the phenomenological scattering rate $\gamma$, which is originally discussed in Ref. [11], Eq. (S12) is expressed as

$$-\omega^2 Q = (B^{11} + i\omega\gamma)Q + B^{12}E. \qquad (S18)$$

By using Eqs. (S13), (S14), and (S18), we derive the system of equations in the presence of the driving force $F_1$ and $F_2$ [12], which are expressed as

$$(-\omega^2 - B^{11} - i\omega\gamma)Q - B^{12}\dfrac{4\pi}{n^2-1}P = F_1, \qquad (S19)$$

$$-B^{21}Q - \dfrac{4\pi B^{22} - n^2 + 1}{n^2-1}P = F_2. \qquad (S20)$$

It has a form $\begin{pmatrix} Q(\omega) \\ P(\omega) \end{pmatrix} = G_{ij}(\omega)\begin{pmatrix} F_1(\omega) \\ F_2(\omega) \end{pmatrix}$, where $G_{ij}(\omega)$ is the Green's function. When $F_1(\omega)$ is much larger than $F_2(\omega)$, $Q(\omega)$ is dominated by $G_{11}(\omega)$ [13], which is given by

$$G_{11}(\omega) = \dfrac{c^2 k^2 - \varepsilon_\infty \omega^2}{\varepsilon_\infty\left(\omega^4 + \omega^3(i\gamma) - \omega^2 \dfrac{c^2 k^2 - \varepsilon_\infty \omega_{TO}^2}{\varepsilon_\infty} - \omega\dfrac{i\gamma c^2 k^2}{\varepsilon_\infty} + \dfrac{\omega_{TO}^2 c^2 k^2}{\varepsilon_\infty}\right)} \qquad (S21),$$

By using Eq. (S21), we can derive

$$Q(\omega) = R_{jk} \frac{c^2k^2 - \varepsilon_\infty \omega^2}{\varepsilon_\infty(\omega^4 + \omega^3(2i\gamma) - \omega^2 \frac{c^2k^2 - \varepsilon_\infty \omega_{TO}^2}{\varepsilon_\infty} - \omega \frac{2i\gamma c^2k^2}{\varepsilon_\infty} + \frac{\omega_{TO}^2 c^2k^2}{\varepsilon_\infty})} E_j(\omega)E_k^*(\omega), \quad (S22)$$

in the main text.

Table SI. Selection-rule of Raman-active modes and mode assignments using *bc*-, *ab*-, and *ac*-surface crystals in the backscattering geometry. Asterisk indicates the mode due to the leakage.

| Configurations | Representations | $A_1(z)$ TO (THz) | $A_1(z)$ LO (THz) | $A_2$ (THz) | $B_1(x)$ (THz) | $B_2(y)$ (THz) |
|---|---|---|---|---|---|---|
| $b(aa)\bar{b}$ | $A_1(z)$ TO | 0.99<br>2.40 | *1.20<br>*2.05 | – | – | – |
| $b(cc)\bar{b}$ | $A_1(z)$ TO | 0.99<br>2.40 | – | – | *0.82 | – |
| $b(ac)\bar{b}$ | $B_1(x)$ | – | – | – | 0.82<br>1.3<br>2.4<br>2.67 | – |
| $b(ca)\bar{b}$ | $B_1(x)$ | – | – | – | 0.82<br>1.3<br>2.4<br>2.67 | – |
| $c(bb)\bar{c}$ | $A_1(z)$ LO | – | 0.99<br>1.21<br>1.80<br>2.06<br>2.38 | – | *0.82 | – |
| $c(aa)\bar{c}$ | $A_1(z)$ LO | – | 0.99 | – | – | – |

|  |  |  | 1.21 |  |  |
|  |  |  | 1.80 |  |  |
|  |  |  | 2.06 |  |  |
|  |  |  | 2.38 |  |  |
| $c(ab)\overline{c}$ | $A_1$ | – | *1.20 | 0.82<br>1.77<br>2.40 | – | – |
| $c(ba)\overline{c}$ | $A_1$ | – | *1.20 | 0.82<br>1.77<br>2.40 | – | – |
| $a(cc)\overline{a}$ | $A_1(z)$ TO | 0.99<br>2.40 | – | *1.70 | – | – |
| $a(bb)\overline{a}$ | $A_1(z)$ TO | 0.98<br>2.40 | *1.20<br>*2.05 | – | – | – |
| $a(cb)\overline{a}$ | $B_2(y)$ | – | – | – | – | 1.80<br>2.30<br>2.67 |
| $a(bc)\overline{a}$ | $B_2(y)$ | – | – | – | – | 1.80<br>2.30<br>2.67 |

# References

[1] A. K. Tagantsev, L. E. Cross, and J. Fousek, *Domains in Ferroic Crystals and Thin Films* (Springer, New York, 2010).

[2] M. Sotome, N. Kida, S. Horiuchi, and H. Okamoto, Visualization of ferroelectric domains in a hydrogen-bonded molecular crystal using emission of terahertz radiation, Appl. Phys. Lett. **105,** 041101 (2014).

[3] M. Sotome, N. Kida, S. Horiuchi, and H. Okamoto, Terahertz radiation imaging of ferroelectric domain topography in room-temperature hydrogen-bonded supramolecular ferroelectrics, ACS Photonics **2,** 1373-1383 (2015).

[4] Y. Kinoshita, N. Kida, M. Sotome, R. Takeda, N. Abe, M. Saito, T. Arima, and H. Okamoto, Visualization of ferroelectric domains in boracite using emission of terahertz radiation, Jpn. J. Appl. Phys. **53,** 09PD08 (2014).

[5] M. Sotome, N. Kida, Y. Kinoshita, H. Yamakawa, T. Miyamoto, H. Mori, and H. Okamoto, Visualization of a nonlinear conducting path in an organic molecular ferroelectric by using emission of terahertz radiation, Phys. Rev. B **95,** 241102(R) (2017).

[6] Y. Kinoshita, N. Kida, M. Sotome, T. Miyamoto, Y. Iguchi, Y. Onose, and H. Okamoto, Terahertz radiation by subpicosecond magnetization modulation in the ferrimagnet $LiFe_5O_8$, ACS Photonics **3,** 1170-1175 (2016).

[7] Y. R. Shen, *The Principles of Nonlinear Optics* (Wiley, New York, 1984).

[8] M. Bass, P. A. Franken, and J. F. Ward, and G. Weinreich, Optical rectification, Phys. Rev. Lett. **9,** 446-448 (1962).

[9] M. Sotome, N. Kida, R. Takeda, and H. Okamoto, Terahertz radiation induced by coherent phonon generation via impulsive stimulated Raman scattering in paratellurite, Phys. Rev. A **90,** 033842 (2014).

[10] C. M. Tu, S. A. Ku, W. C. Chu, C. W. Luo, J. C. Chen, and C. C. Chi, Pulsed terahertz radiation due to coherent phonon-polariton excitation in <110> ZnTe crystal, J. Appl. Phys. **112,** 093110 (2012).

[11] R. Takeda, N. Kida, M. Sotome, Y. Matsui, and H. Okamoto, Circularly polarized terahertz radiation from a eulytite oxide by a pair of femtosecond laser pulses, Phys. Rev. A **89,** 033832 (2014).

[12] R. Claus, L. Merten, and J. Brandmüller, *Light Scattering by Phonon-Polaritons* (Springer-Verlag, Berlin, 1975).

[13] T. P. Dougherty, G. P. Wiederrecht, and K. A. Nelson, Impulsive stimulated Raman scattering experiments in the polariton regime, J. Opt. Soc. Am. B **9**, 2179-2189 (1992).

[14] K. Huang, Lattice vibrations and optical waves in ionic crystals, Nature (London) **167**, 779-780 (1951).